\documentclass[conference]{IEEEtran}
\IEEEoverridecommandlockouts
\usepackage{cite}
\usepackage[normalem]{ulem} 
\usepackage{amsmath,amssymb,amsfonts}
\usepackage{algorithmic}
\usepackage{graphicx}
\usepackage{textcomp}
\usepackage{xcolor}
\usepackage{quantikz,adjustbox}
\def\BibTeX{{\rm B\kern-.05em{\sc i\kern-.025em b}\kern-.08em
    T\kern-.1667em\lower.7ex\hbox{E}\kern-.125emX}}

\usepackage{acronym}

\acrodef{PDE}[PDE]{Partial Differential Equation}
\acrodef{IWE}[IWE]{Isotropic Wave Equation}
\acrodef{QC}[QC]{Quantum Computer}
\acrodef{QCs}[QCs]{Quantum Computers}
\acrodef{VQA}[VQA]{Variational Quantum Algorithm}
\acrodef{VQAs}[VQAs]{Variational Quantum Algorithms}
\acrodef{MPSs}[MPSs]{Matrix Product States}
\acrodef{MPS}[MPS]{Matrix Product State}
\acrodef{MPO}[MPO]{Matrix Product Operator}
\acrodef{MPOs}[MPOs]{Matrix Product Operators}

\begin{document}

\title{A Quantum-Inspired Algorithm for Wave Simulation Using Tensor Networks
}

\author{
\IEEEauthorblockN{Kevin Lively}
\IEEEauthorblockA{\textit{Institute of Software Technology} \\
\textit{German Aerospace Center (DLR)}\\
Sankt Augustin, Germany \\
ORCID: 0000-0003-2098-1494 }
\and
\IEEEauthorblockN{Vittorio Pagni}
\IEEEauthorblockA{\textit{Institute of Software Technology} \\
\textit{German Aerospace Center (DLR)}\\
Sankt Augustin, Germany \\
\textit{University of Cologne}\\
Cologne, Germany\\
ORCID: 0009-0006-9753-3656}\\
\and
\IEEEauthorblockN{Gonzalo Camacho}
\IEEEauthorblockA{\textit{Institute of Software Technology} \\
\textit{German Aerospace Center (DLR)}\\
Sankt Augustin, Germany \\
ORCID: 0000-0001-6900-8850 }
}

\maketitle

\begin{abstract}
    We present an efficient classical algorithm based on the construction of a unitary quantum circuit for simulating the \ac{IWE} in one, two, or three dimensions. Using an analogy with the massless Dirac equation, second order time and space derivatives in the \ac{IWE} are reduced to first order, resulting in a Schrödinger equation of motion. Exact diagonalization of the unitary circuit in combination with Tensor Networks allows simulation of the wave equation with a resolution of $10^{13}$ grid points on a laptop. A method for encoding arbitrary analytical functions into diagonal \ac{MPOs} is employed to prepare and evolve a \ac{MPS} encoding the solution. Since the method relies on the Quantum Fourier Transform, which has been shown to generate small entanglement when applied to arbitrary \ac{MPS}s, simulating the evolution of initial conditions with sufficiently low bond dimensions to high accuracy becomes highly efficient, up to the cost of Trotterized propagation and sampling of the wavefunction. We conclude by discussing possible extensions of the approach for carrying out Tensor Network simulations of other partial differential equations such as Maxwell's equations.
\end{abstract}


\section{Introduction}
Due to their theoretical capacity to manipulate data of size $2^n$ within the Hilbert space of $n$ qubits, \ac{QCs}  have been advanced as promising platforms for large scale simulation of Partial Differential Equations (PDEs). However, the physics of quantum devices is ultimately governed by a single \ac{PDE}, namely the Schrödinger equation. Thus, mapping other \ac{PDE}s into forms which can be simulated by the evolution of a quantum mechanical wavefunction is the first challenge in designing any Quantum Algorithm for this task \cite{Berry2014,Costa2019,Vahala2020,Gaitan2020,Childs2020,An2023,Koukoutsis2023,Koukoutsis2023App,Vahala2023,Fang2023,Brearley2024,Wright2024,Jin2024,Jin2024_2,Jin2023,Hu2024quantumcircuits,Sato2025,Tennie2025}. Subsequently, dozens of approaches to this problem have been put forward in the literature spanning both hybrid quantum-classical and fully-quantum algorithms. Many of these approaches are either based on linearizing the PDE over time and space derivatives along with a Carleman approach for sufficiently weak non-linear terms, then either inverting the resulting system of equations using the Harrow-Hassidim-Lloyed algorithm \cite{HHL,Childs2017, Clader2013,Berry2014,Wang2020,Childs2020,Liu2021}, constructing a variational quantum algorithm  \cite{Lubasch2020,Pool2022,Jaksch2023,Pool2024}, or using QCs to (theoretically) speed up bottleneck integrations in integro-differential approaches \cite{Gaitan2020,Gaitan2021,Oz2021,Oz2023}. A crucial aspect to this field of research is that while many of these algorithms appear to be impractical for near-term hardware, some of them have inspired highly efficient classical methods \cite{Lubasch2018,Lubasch2020,GarciaRipoll2021quantuminspired,Gourianov2022,Jaksch2023,Kiffner2023,Peddinti2024}. Given the centrality of PDE simulation across virtually every field of science, technology and engineering, fresh perspectives on this problem can have a substantial and immediate impact, as has been recently seen in novel computational fluid dynamics methods \cite{Gourianov2022,Kiffner2023,Peddinti2024,Gourianov2025}.

The wave equation is a \ac{PDE}, whose simulation spans critical application areas in fluid dynamics, electromagnetism, seismology and medicine \cite{Sahin2023,Lyu2022,Okita2022,Chew2009}. Numerical approaches for modeling wave propagation are generally grouped into real-space finite differencing methods or spectral approaches. Due to second-order derivatives in space and time, real space methods typically require high order stencils on a fine grid and tracking of multiple time steps simultaneously to be stable and accurate \cite{Etgen2007}. In contrast, spectral methods require conversion between momentum-space and real-space via the Fast Fourier Transform (FFT), or analogous polynomial decompositions, which for the usually large grid sizes involved in three-dimensional (3D) simulation can be difficult to parallelize, due to requiring an all-to-all operation \cite{Ma2021}. In the limit of a uniform medium, the problem reduces to the Isotropic Wave Equation (IWE), an exactly solvable \ac{PDE} which is often used as an introductory example to build up more complicated \ac{PDE} simulation techniques. Thus this serves as a natural starting point when approaching the problem of simulating \ac{PDE}s employing quantum algorithms.

In a recent work by Wright et. al. a quantized representation of the IWE in one dimension (1D) was developed~\cite{Wright2024}, followed by implementing a concrete unitary circuit based on the abstract approach of Costa, Jordan and Ostrander \cite{Costa2019}. By relying on the Quantum Fourier Transform (QFT) to map to reciprocal space, they found a unitary circuit capable of simulating the 1D IWE in the small wavenumber limit to arbitrary times in a fixed depth and demonstrated it on Quantinuum's Ion Trap hardware. Two notable aspects of this work stand out for our purposes. One is that the central reliance on the QFT provides an appealing opportunity for quantum-inspired algorithms. Recently it was rigorously shown that the core circuit in the QFT, i.e. the combination of Hadamard and controlled phase gates without the bit reversal operation, can be very efficiently implemented using Tensor Networks \cite{Chen2023}. Specifically using the Matrix Product State (MPS) or Tensor Train structure \cite{Vidal2003,Verstraete2008,Oseledets2011,Schollwoeck2011,Orus2014}, the resultant Matrix Product Operator (MPO) representation of the core of the QFT can be constructed to a high degree of accuracy with a fixed cost in the bond dimension, allowing it to outperform the FFT on data which can be efficiently represented as MPSs. The second notable aspect is that the method of Costa et. al relies on constructing a Hamiltonian operator whose square, i.e the second time derivative of the wavefunction, returns the Laplacian. This suggests a direct analogy with the Dirac equation in Quantum Mechanics, which we explicitly exploit in order to develop the associated quantum representation of the original problem. In this work, we combine these two insights to develop an MPS based algorithm capable of simulating the IAW in one, two or three dimensions. 

The paper is structured as follows. In Section II we lay out our derivation based on the Dirac Equation, whose square also returns the IAW in the massless limit. In section III, using a unitary circuit representation of the equation of motion in reciprocal space, we derive an exact solution and benchmark the results against direct integration of the real space equations of motion. We also describe how to approximately propagate the wave equation via Trotterization. In section IV we utilize an encoding method based on the sum of Pauli operators containing only $\hat{I}$ and $\hat{Z}$ operators in order to prepare simple analytical functions onto the diagonals of MPOs without having to actually hold them in memory, and use this to prepare such functions as MPSs. As with other methods for representing functions in a tensor network format \cite{Oseledets2010,Oseledets2013,Keller2015,Ali2023,Melnikov2023,GonzalezConde2024efficientquantum} this allows us to push the effective grid size we can simulate well beyond classical storage limits, which we demonstrate by encoding a two dimensional Ricker wavelett into a $100$ qubit site MPS, equivalent to about $10^{30}$ grid points. We connect our results to the work of Wright et. al by showing how, in the small wavenumber limit, our method reduces to theirs in one dimension. In section V we show the scaling of our MPS dynamics in the case of a two-dimensional Ricker wavelett, comparing it with a direct evolution using the FFT and our reciprocal space exact solution. We find that even a very simplistic MPS evolution scheme obtains advantage in computational time on grids of about $10^{9}$ points, mainly due to the classical computational advantage of the MPO-QFT, with a maximum absolute error rate of $0.86\%$. We also consider an approximate set of initial conditions for three dimensions in order to analyze the numerical efficiency of our method, again finding a speed up around $10^{9}$ grid points due to the MPO-QFT. We conclude with a discussion of the implications of this work and the potential future research directions. 

\section{Derivation of the \ac{IWE} as a Schrödinger Equation}
\subsection{The Dirac Equation}
For the sake of being self contained we will be explicit in the details of deriving Dirac's equation and how it connects to the IWE. We first begin with the Acoustic \ac{IWE} in three dimensions, setting the velocity of sound $c=1$:
\begin{equation}\label{IWE}
    (\partial_t^2 - \partial_x^2 - \partial_y^2 -\partial_z^2)u(t,x,y,z) = 0.
\end{equation}
We will find it convenient to adopt the notation normally used when working in the Minkowski metric:
\begin{equation}
    \partial_{\mu} = (\partial_t,\partial_x,\partial_y,\partial_z)
\end{equation}
\begin{equation}
    \partial^{\mu} = (\partial_t,-\partial_x,-\partial_y,-\partial_z)
\end{equation}
\begin{equation}
\begin{split}
    \partial^{\mu}\partial_{\mu} &= \partial_t^2 - \boldsymbol{\nabla}\cdot\boldsymbol{\nabla}\\
    &= \partial_t^2 - (\partial_x^2 + \partial_y^2 +\partial_z^2).
\end{split}
\end{equation}
We use Einstein notation throughout much of this text, i.e. when an index $\mu\in\{t,x,y,z\}$ is repeated in both a raised and lowered position, it is summed over, as above.
Eq. \eqref{IWE} is of course second order in its time and space derivatives. To be able to write it in a Schrödinger like equation, i.e. first order in time derivatives, it would be helpful to define the object $\sqrt{\partial^{\mu}\partial_{\mu}} = \sqrt{\partial^2}$. Of course this is the same problem faced by Dirac when dealing with inadequacies of the Klein-Gordon equation in Quantum Field Theory \cite{Lancaster2014}:
\begin{equation}
    (\partial^2+m^2)\psi = 0.
\end{equation}
In this paper we are dealing with the massless limit, $m=0$. The solution is to define four $4\times 4$ matrices each labeled by $\mu$: $\gamma^{\mu}$. This collection of matrices obeys the Clifford Algebra:
\begin{equation}
    (\gamma^{\mu})^2 = 1\null\qquad \{\gamma^{\mu},\gamma^{\nu}\}=2\eta^{\mu\nu}, 
\end{equation}
where $\{\gamma^{\mu},\gamma^{\nu}\}$ is the anti-commutator and $\eta^{\mu\nu}=\text{diag}(1,-1,-1,-1)$ is the Minkowski metric tensor in the mostly-minus convention. We adopt the Weyl representation for the $\gamma$ matrices:
\begin{equation}\label{gamma-matrices}
    \gamma^t = \begin{pmatrix}
    0 & I_{2\times 2} \\
    I_{2\times 2} & 0 
    \end{pmatrix},\null\quad \gamma^l = \begin{pmatrix}
    0 & \sigma^l \\
    -\sigma^l & 0
    \end{pmatrix},
\end{equation}
where we use the convention that greek indices $\mu,\nu\in\{t,x,y,z\}$ correspond to space and time indices and latin indices $l,m\in\{x,y,z\}$ correspond to exclusively the spatial components. The $\sigma^l$ matrices correspond to the normal $X, Y, Z$ Pauli matrices, and $I_{n\times n}$ is the $n\times n$ identity matrix. It is then straightforward to show that
\begin{equation}
    (\gamma^\mu\partial_{\mu})^2 = I_{4\times 4}\partial^2.
\end{equation}
The introduction of the $\gamma^{\mu}$ matrices imposes additional structure on our differential equation: we are now working with $4\times 1$ vectors subject to the Clifford Algebra, referred to as spinors. From Eq. \eqref{gamma-matrices} we see that in our massless case the set of four differential equations splits into two sectors, which we call ``left'' and ``right''. We thus write our spinor wavefunction as $\Psi(t,x,y,z) = (\Psi_L,\Psi_R)^T$ where the  $\Psi_{L/R}\coloneqq\Psi_{L/R}(t,x,y,z)$ elements are themselves $2\times 1$ vectors with scalar complex functions as entries. Thus our system of equations becomes
\begin{equation}
     (\gamma^{\mu}\partial_{\mu})\Psi = (\gamma^t\partial_t - \boldsymbol{\gamma}\cdot\boldsymbol{\nabla})\Psi = 0,
\end{equation}
equivalent to
\begin{equation}\label{DiracSpinorEOM}
    \begin{split}
        I_{2\times 2}\partial_t\Psi_R &= \boldsymbol{\sigma}\cdot\boldsymbol{\nabla}\Psi_R \\
        I_{2\times 2}\partial_t\Psi_L &= -\boldsymbol{\sigma}\cdot\boldsymbol{\nabla}\Psi_L.\\
    \end{split}
\end{equation}
For clarity we suppress the dependence on $(t,x,y,z)$. We will see shortly that the left and right wavefunctions correspond to forward and backward propagating solutions to the IWE. For the moment though we inspect what the two components of the $L/R$ vector correspond to. Take $\Psi_R = (\psi_0,\psi_1)$. The coupled equations are then:
\begin{equation}\label{eq: direct EOM}
\begin{split}
    \partial_t\psi_0 &= \partial_z\psi_0 + (\partial_x-i\partial_y)\psi_1 \\
    \partial_t\psi_1 &= (\partial_x+i\partial_y)\psi_0 - \partial_z\psi_1. \\
\end{split}
\end{equation}
Taking the second time derivative of $\psi_0$ gives us
\begin{equation}
\begin{split}
    \partial_t^2\psi_0 = (\partial_x^2 + \partial_y^2 + \partial_z^2)\psi_0 
\end{split}
\end{equation}
thereby recovering the IWE. So if we allow $\psi_0(0,x,y,z)$ to represent the initial spatial conditions of our IWE, and specify $\partial_t\psi_0(0,x,y,z)$ we can initialize $\psi_1(0,x,y,z)$ as
\begin{equation}
    \psi_1(0,x,y,z) = (\partial_x-i\partial_y)^{-1}(\partial_t\ - \partial_z)\psi_0(0,x,y,z). 
\end{equation}
For the moment we leave the inverse derivative operator as an abstract symbol. With appropriate choices of boundary conditions or numerical approximation this inverse operator can be constructed to solve for $\psi_1$. From the appearance of the partial derivative in time, we see that $\psi_1$ is related to the initial conditions of our wavepacket.

\subsection{Encoding the Real-Space Dirac Equation onto Qubits}
Now, we allow $\psi_0$ and $\psi_1$ to be elements of a Hilbert space across the spatial directions, $\mathcal{H}_r = \mathcal{H}_x\otimes\mathcal{H}_y\otimes\mathcal{H}_z$. In this perspective the derivative operators act on their respective Hilbert spaces, $\hat{\partial}_l:\mathcal{H}_l\to\mathcal{H}_l$. To simplify notation we will write $\hat{I}_x\otimes\hat{\partial}_y\otimes \hat{I}_z = \hat{\partial}_y$ with the identity operators on complementary subspaces $\hat{I}_x/\hat{I}_z$ being taken for granted.

In this formulation, we can take the $\Psi_R$ equation of motion  (EOM) from Eq. \eqref{DiracSpinorEOM}. Now since this (up to a global phase) is equivalent to the EOM for $\Psi_L$, we may as well drop this $L/R$ for the time being and continue focusing on the $2\times 1$ vector $\Psi$. Thus we write:
\begin{equation}\label{waveEq-space}
    \hat{\partial}_t\ket{\Psi} =  (\hat{X}\otimes \hat{\partial}_x + \hat{Y}\otimes \hat{\partial}_y + \hat{Z}\otimes\hat{\partial}_z)\ket{\Psi}.
\end{equation}
The total Hilbert space in which $\ket{\Psi}$ exists is $\mathcal{H} = \mathcal{H}_0\otimes\mathcal{H}_r$:  the Pauli operators $\hat{X},\hat{Y},\hat{Z}$ act on the single qubit space $\mathcal{H}_0$, and this is the same qubit space which assigns $\psi_0(x,y,z)$ to the first entry of the $2\times 1$ spinor and $\psi_1(x,y,z)$ to the second entry. We will assign it as the left-most digit in the register, and name it the ``time'' qubit, although it's really responsible for coupling $\psi_0$ and $\psi_1$.

Again for the sake of self-containment we are explicit in our definition of the real space and reciprocal space bases.
We represent $\psi_{0/1}(x,y,z)$ on a finite grid of $N_l=2^{n_l}$ points in each direction $l$ as $\psi_{0/1}(x_i,y_j,z_k)$. In this context, $i,j,k\in\{0,N-1\}$ are integers. Each $x_i=i\Delta x$  corresponds to a point on an even, unit length subdivision, i.e $\Delta x = \Delta y = \Delta z =1/N$, implicitly defining the simulation box length to be $1$. On this grid we can approximately resolve $\hat{\partial}_l$ using a central finite difference operator. For example in the $x$ direction:
\begin{equation}
    \big(\partial_x\psi(x_i,y_i,z_i)\big)\approx \frac{1}{2\Delta x}\big( \psi(x_{i+1},y_j,z_k) - \psi(x_{i-1},y_j,z_k) \big).
\end{equation}
This approximation becomes exact in the limit of $\Delta x\to0$. We adopt Periodic Boundary Conditions (PBC) in all directions such that $x_{i+N} = x_i$. The position vector-operator $\hat{\mathbf{r}} = (\hat{x},\hat{y},\hat{z})$ has the obvious unit-step eigenvectors and eigenvalues:
\begin{equation}
    (\hat{x},\hat{y},\hat{z}) \ket{x_i,y_i,z_i}=(x_i,y_i,z_i)\ket{x_i,y_i,z_i}.
\end{equation} On this periodic grid the operator $\hat{\partial}_l$ can be diagonalized by the QFT when it acts on the $\mathcal{H}_l$ subspace. We label each subspace's QFT as $\hat{Q}_l$:
\begin{equation}
    \hat{\partial}_l = -i\hat{Q}_l^{\dagger}\hat{D}_l\hat{Q}_{l}.
\end{equation}
where the entries of the diagonal matrix $\langle k_m|\hat{D}_l|k_n\rangle=p_n\delta_{mn}$ are proportional to the eigenvalues of the central finite difference operator
\begin{equation}
\begin{split}
    \hat{D}_l &= N\text{sin}(2\pi k_l/N)|k_l\rangle\langle k_l|\\
    &= p_l|k_l\rangle\langle k_l|.
\end{split}
\end{equation}
Here $k_l\in\{-N_l/2,\ldots N_l/2-1\}$ are integers since $N_l=2^{n_l}$, and we refer to $\ket{k_l}$ as the wavenumber basis, which diagonalize the momentum operator. These are planewaves in the real-space basis:
\begin{equation}
    \langle\mathbf{k}|\mathbf{r}\rangle = e^{-i\mathbf{k}\cdot\mathbf{r}}.
\end{equation}
By taking the product of the QFT across each subspace, $\hat{Q}=\hat{Q}_x\otimes \hat{Q}_y\otimes \hat{Q}_z$ and applying it to Eq. \eqref{waveEq-space} we obtain
\begin{equation}\label{SchrödingerEq}
    \hat{\partial}_t\ket{\Psi} = -i \hat{Q}^{\dagger}( \hat{X}\otimes \hat{D}_x + \hat{Y}\otimes \hat{D}_y + \hat{Z}\otimes \hat{D}_z )\hat{Q}\ket{\Psi},
\end{equation}
Thereby encoding the evolution \ac{IWE} in $\Psi$. This is the time-independent Schrödinger equation. When measuring the time-qubit we collapse the spatial qubit state to be either $\psi_0$ or $\psi_1$ when we measure $0$ or $1$ respectively. It is straightforward to see that had we taken the $\Psi_L$ equation above, we would achieve the same result but with a a global phase giving a positive $i$. We infer then that $\Psi_L$ then gives the backwards propagating solution. $t\to -t$.

\section{Exact and Approximate Propagation}
\subsection{Exact Propagation}
In this section we find an efficient representation of the time evolution operator in the QFT transformed basis. From Eq. \eqref{SchrödingerEq}, we see that the exact expression for our wavefunction at time $t$ in terms of our initial state $\ket{\Psi_0}$ is
\begin{equation}\label{eq: evolution eq.}
\begin{split}
    \ket{\Psi(t)} &= e^{-i\hat{H}t}\ket{\Psi_0}\\
    &=\hat{Q}^{\dagger}e^{-i\left(\hat{X}\otimes \hat{D}_x + \hat{Y}\otimes \hat{D}_y + \hat{Z}\otimes  \hat{D}_z\right)t}\hat{Q}\ket{\Psi_0},
\end{split}
\end{equation}
where we have used the fact that 
\begin{equation}
    e^{-it\hat{\sigma}^i\otimes(\hat{Q}^{\dagger}\hat{D}_i\hat{Q})} = \hat{Q}^{\dagger}e^{-it\hat{\sigma}^i\otimes \hat{D}_i}\hat{Q}.
\end{equation}
This means that our problem ultimately boils down to finding a useful expression for the QFT transformed Hamiltonian
\begin{equation}
    \begin{split}
        \hat{H}&=\hat{X}\otimes \hat{D}_x + \hat{Y}\otimes \hat{D}_y + \hat{Z}\otimes  \hat{D}_z.\\
    \end{split}
\end{equation}
Note that we have chosen the order of application of $\hat{Q}$ and $\hat{Q}^{\dagger}$ very deliberately. Recall that the QFT is defined with a complete bit reversal operation performed by pairwise SWAP operations across the quantum register as the last step. To avoid cluttering the notation, here do not yet consider the effects of conjugating the diagonal $\hat{D}_l$ operators by the bit reversal operation, so that we can simply study these operators. This will be addressed in the following section. 
Resolving this in our wavevector basis we have
\begin{equation}
\begin{split}
    \hat{H} &= \sum_{k_xk_yk_z} \bigg(\hat{\sigma}^{x}p_x(k_x) +\hat{\sigma}^{y}p_y(k_y) +\hat{\sigma}^{z}p_z(k_z)\bigg) \otimes\ket{\mathbf{k}}\bra{\mathbf{k}}\\
    &=\sum_{\mathbf{k}} |\mathbf{p}|\hat{h}(\mathbf{k})\otimes\ket{\mathbf{k}}\bra{\mathbf{k}}.
\end{split}
\end{equation}
Where
\begin{equation}
\begin{split}
    \hat{h}(\mathbf{k}) &= \frac{1}{|\mathbf{p}|}\sum_l\hat{\sigma}^lp_l(k_l),
\end{split}
\end{equation}
$\ket{\mathbf{k}}\coloneqq\ket{k_xk_yk_z}$, $|\mathbf{p}|\coloneqq\sqrt{p_x^2+p_y^2+p_z^2}$, and we have reemphasized that the momenta eigenvalues $p_l$ are functions of the wavevectors $k_l$. We see that this is a block-diagonal matrix, namely the wavevector states don't interact with each other. Thus we can think about the problem in terms of various points in wavevector-momenta space. Having $\hat{h}(\mathbf{k})$ normalized with respect to $|\mathbf{p}|$ allows us to think in terms of spherical coordinates:
\begin{equation}
    \begin{split}
        p_x &= |\mathbf{p}|\text{sin}(\theta_z)\text{cos}(\phi_{x,y})\\
p_y &= |\mathbf{p}|\text{sin}(\theta_z)\text{sin}(\phi_{x,y})\\
p_z &= |\mathbf{p}|\text{cos}(\theta_z).
    \end{split}
\end{equation}
where $\theta_z=\text{arccos}(p_z/|\mathbf{p}|)$ is the angle from the $p_z$ axis, and $\phi_{x,y} = \text{arctan}(p_y/p_x)$ is the angle from the $p_x$ axis in the $p_x-p_y$ plane. Of course, starting from a rectilinear grid and moving to spherical coordinates requires careful handling of the edge cases for $\phi_{x,y}$ and $\theta_z$, but we write arccos and arctan with this understood. In this perspective then, each $\mathbf{k}$ block's kernel hamiltonian $\hat{h}(\mathbf{k})$ can be written in the $Z$ computational basis as
\begin{equation}
\begin{split}
h(\mathbf{k}) &=
\frac{1}{|\mathbf{p}|}\begin{pmatrix}
p_z &p_x-ip_y\\
p_x+ip_y&-p_z
\end{pmatrix}\\
&=
\begin{pmatrix}
    \text{cos}(\theta_z) &\hspace{-0.2cm} \text{sin}(\theta_z)e^{-i\phi_{x,y}}\\
    \text{sin}(\theta_z)e^{i\phi_{x,y}} &\hspace{-0.2cm} -\text{cos}(\theta_z)\\
    \end{pmatrix}.
\end{split}
\end{equation}
This kernel can be diagonalized by eigenvectors
\begin{equation}
    \begin{split}
        \chi_+ &= \begin{pmatrix}
            \text{cos}(\frac{\theta_{z}}{2})\\
         e^{i\phi_{x,y}}\text{sin}(\frac{\theta_{z}}{2})
        \end{pmatrix}\\
        \chi_- &= \begin{pmatrix}
        -\text{sin}(\frac{\theta_{z}}{2})\\
         e^{i\phi_{x,y}}\text{cos}(\frac{\theta_{z}}{2})\\
        \end{pmatrix}\\
    \end{split}
\end{equation}
with associated eigenvalues $\hat{h}(\mathbf{k})\ket{\chi_{\pm}}=\pm\ket{\chi_{\pm}}$. Putting all this together means that
\begin{equation}\label{eq:eq_expH}
\begin{split}
e^{-it\hat{H}}&=\sum_{\mathbf{k}}e^{-i|\mathbf{p}|\hat{h}(\mathbf{k}) t}\otimes|\mathbf{k}\rangle\langle \mathbf{k}|\\
&=\sum_{\mathbf{k}}\bigg(e^{-i|\mathbf{p}|t}\ket{\chi_+}\bra{\chi_+} + e^{i|\mathbf{p}|t}\ket{\chi_-}\bra{\chi_-}\bigg)\otimes|\mathbf{k}\rangle\langle \mathbf{k}|\\
&=\sum_{\mathbf{k}}V^{\dagger}(\mathbf{k})\begin{pmatrix}
e^{-i|\mathbf{p}|t} &0\\
0 & e^{+i|\mathbf{p}|t} 
\end{pmatrix}V(\mathbf{k})\otimes|\mathbf{k}\rangle\langle \mathbf{k}|,
\end{split}
\end{equation}
where $V(\mathbf{k})$ is the change of basis matrix whose columns are the two eigenvectors, $V(\mathbf{k})\coloneqq(\chi_+\ \chi_-)$.

In Eq.~\eqref{eq:eq_expH} we have obtained a diagonal unitary matrix which allows us to evolve any input waveform to arbitrary times, given a $\mathbf{k}$ space representation. We define the wavevector elements of $\ket{\Psi}$ by $\langle\mathbf{k}|\psi_{0/1}\rangle = \psi_{0/1}(\mathbf{k})$. They are obtained from the real space representation by applying $\hat{Q}$. Multiplying $\ket{\Psi}=(\ket{\psi_0},\ket{\psi_1})^{T}$ through Eq.~\eqref{eq: evolution eq.}, and writing the explicit $\mathbf{k}$ momentum dependence, the final form of $\psi_0(t,\mathbf{k})$ and $\psi_1(t,\mathbf{k})$ become
\begin{equation}\label{eq: exact evolution}
\begin{split}
    \psi_0(t,\mathbf{k}) &=
    \bigg[\text{cos}(|\mathbf{p}(\mathbf{k})|t)-i\text{cos}(\theta_z)\text{sin}(|\mathbf{p}(\mathbf{k})|t)\bigg]\psi_0(0,\mathbf{k}) \\
    &- \bigg[ie^{-i\phi_{x,y}}\text{sin}(\theta_z)\text{sin}(|\mathbf{p}(\mathbf{k})|t)\bigg]\psi_1(0,\mathbf{k})\\
    \psi_1(t,\mathbf{k}) &=
    \bigg[-i\text{sin}(|\mathbf{p}(\mathbf{k})|t)\text{sin}(\theta_z)e^{i\phi_{x,y}}\bigg]\psi_0(0,\mathbf{k}) \\
    &+ \bigg[\text{cos}(|\mathbf{p}(\mathbf{k})|t)+i\text{cos}(\theta_z)\text{sin}(|\mathbf{p}(\mathbf{k})|t)\bigg]\psi_1(0,\mathbf{k})
\end{split}
\end{equation}
Subsequently the real-space picture is obtained by back-transformation via $\hat{Q}^{\dagger}$. Thus our exact evolution equation is given by Eq. \eqref{eq: exact evolution}, which is diagonal in wave-vector space, which one would expect given the translational symmetry of the equation.

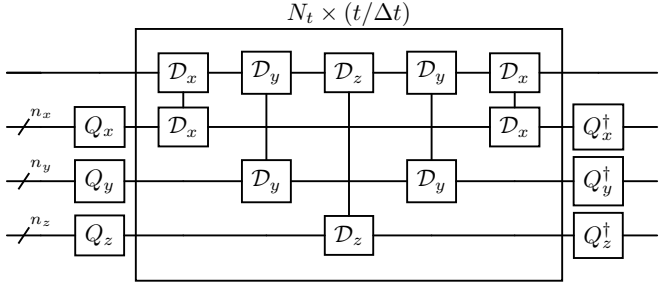
\begin{figure}[h]
    \centering
    \begin{adjustbox}{width=1\linewidth}
    \begin{quantikz}[transparent, row sep={0.8cm,between origins}]
 &  \qw &   & \gate{\mathcal{D}_x}\wire[d][1]{q}\gategroup[4,steps=5,style={inner
sep=6pt}]{$N_t \times (t/\Delta t)$}  &  \gate{\mathcal{D}_y}\wire[d][2]{q} & \gate{\mathcal{D}_z}\wire[d][3]{q}& \gate{\mathcal{D}_y}\wire[d][2]{q} & \gate{\mathcal{D}_x}\wire[d][1]{q} &  & \\
& \qwbundle{n_x} & \gate[1]{Q_x} & \gate{\mathcal{D}_x} & & &  & \gate{\mathcal{D}_x} & \gate[1]{Q^{\dagger}_x} &\\
& \qwbundle{n_y} & \gate[1]{Q_y} &  &\gate{\mathcal{D}_y} & & \gate{\mathcal{D}_y} &   & \gate[1]{Q^{\dagger}_y}&\\
& \qwbundle{n_z} & \gate[1]{Q_z} &  & & \gate{\mathcal{D}_z} & & &  \gate[1]{Q^{\dagger}_z} &
\end{quantikz}
\end{adjustbox}
\caption{Diagram representing either a quantum circuit or an MPO for Eq. \eqref{eq: Trotter MPS Evol} in three spatial directions with $n_x,n_y,$ and $n_z$ qubits each. Here $\mathcal{D}_l = e^{-ic_l\Delta t\hat{\sigma}^l\hat{D}_l}$ and $Q_{l}$ is the Quantum Fourier Transform in the $l^{\text{th}}$ spatial direction. In two dimensions, one can simply drop the $Z$ direction and merge the two $\mathcal{D}_y$ operators. For $X$ and $Y$ $c_l = 1/2$ and for $Z$, $c_l = 1$. In one dimension for the small angle approximation this reduces to the same approach by Wright et. al \cite{Wright2024}, see the text for details. }
\label{fig:Prop full}
\end{figure}

\subsection{Approximate Propagation}
Using classical computation, if one has sufficient memory, calulating Eq. \eqref{eq: exact evolution} is of course straight-forward, requiring only a element-wise vector multiplication. However, as we will see below, representing these functions as MPOs is not so simple. The same holds true for Quantum Circuit construction, where it is not obvious how to encode the unitary evolution equation efficiently. Thus we turn back to the original evolution equation, Eq. \eqref{eq: evolution eq.}. Clearly the three spatial directions do not commute, due to the differing $\hat{\sigma}^l$ operators on the time qubit. However, given some finite timestep $\Delta t$ we can simply Trotterize the evolution in each direction: 
\begin{equation}\label{eq: Trotter MPS Evol}
\begin{split}
    &\hat{U} = e^{-i\frac{\Delta t}{2}\hat{X}\hat{D}_x}e^{-i\frac{\Delta t}{2}\hat{Y}\hat{D}_y}e^{-i\Delta t\hat{Z}\hat{D}_z}e^{-i\frac{\Delta t}{2}\hat{Y}\hat{D}_y}e^{-i\frac{\Delta t}{2}\hat{X}\hat{D}_x}\\
    &\ket{\Psi(t)}\approx \hat{Q}^{\dagger} (\hat{U})^{N_t}\hat{Q}\ket{\Psi_0},
\end{split}
\end{equation}
where $N_t = t_f/\Delta_t$ is the number of timesteps required to get to the desired final time $t_f$. This is the simplest form of Trotterization for three non-commuting terms and will generally scale with error $\mathcal{O}(\Delta t^2)$, although more advanced splitting will produce better error scaling \cite{Barthel2020}. A schematic of this abstract circuit or equivalent MPO is seen in Fig \ref{fig:Prop full}. Now we must turn to the question of how to actually encode our initial states, and construct these operators.

\section{State and Operator Encoding}
For an $N$ qubit system and the associated computational basis states $\{|\sigma_1...\sigma_N\rangle\}$ with $\sigma_l\in\{0,1\}$, generic quantum states represented by MPS and operators represented by MPO have forms:
\begin{eqnarray}
\text{MPS} &\to& \sum_{\{\sigma_l\}}M^{\sigma_1}...M^{\sigma_N}|\sigma_1,...,\sigma_N\rangle,\nonumber\\
\text{MPO} &\to& \sum_{\{\sigma_l\},\{\sigma'_l\}}W^{\sigma_1,\sigma'_1}...W^{\sigma_N,\sigma'_N}|\sigma_1...\sigma_N\rangle\langle \sigma'_1...\sigma'_N|,\nonumber\\
\end{eqnarray}
where for fixed value of $\sigma_l$, $M^{\sigma_l}$ are matrices of dimension $(\chi_l,\chi_{l+1})$, whereas for fixed $\sigma_l,\sigma'_l$ $W^{\sigma_l,\sigma'_l}$ is a matrix of dimension $(b_l,b_{l+1})$. The quantities $\chi_l,b_l$ represent the corresponding bond dimensions for the MPS and MPO, respectively. These tensors can be written in a graphical manner, whose structure is essentially the same as Quantum Circuit notation, as in Fig. \ref{fig:Prop full}. Out of space considerations we defer to the standard literature for more details~\cite{Verstraete2008,Oseledets2011,Schollwoeck2011}. 

\subsection{MPS and MPO encoding}

Let $q_i\in\{1,\ldots,n_l\}$ correspond to the $i^{\text{th}}$ qubit in the set of $n_l$ qubits responsible for the $N_l=2^{n_l}$ grid points in spatial direction $l$. Recall that for example in the x direction $x_i = i\Delta x $, where $i\in\{0,\ldots,2^{n_x}-1\}$. We can define the ``index operator'' $\hat{\imath}$ such that $\hat{\imath}\ket{i} = i\ket{i}$, and $\langle i|j\rangle=\delta_{ij}$. Each ket here is the integer represented in binary, i.e. $\ket{0}=\ket{0}^{\otimes n_x},\ \ket{1} = \ket{0}^{\otimes n_x-1}\ket{1},\ldots ,\ \ket{2^{n_x}-1} = \ket{1}^{\otimes n_x}$. In other words, in the computational basis this operator is a diagonal matrix with entries $i\in\{0,\ldots,2^{n_x}-1\}$. It turns out we can write the index operator in terms of Pauli operators as such:
\begin{equation}\label{eq: index operator}
\begin{split}
    \hat{\imath} &= \sum_{i=0}^{2^{n_x}-1}i\ket{i}\bra{i}\\
    &=\sum_{q_i=1}^{n_x} 2^{n_x-q_i-1}(\hat{I}^{\otimes n_x} - \hat{I}\otimes\ldots \otimes \hat{Z}_{q_i}\otimes\ldots \hat{I}),
\end{split}
\end{equation}
where the second term on the bottom line is simply the Pauli with $\hat{Z}$ on qubit $q_i$ and $\hat{I}$ everywhere else. This expression, which encodes the index along the diagonal of an operator, turns out to be very powerful as one can directly express any function over the index as a function of the index operator. For example, take the Ricker Wavelett function centered at $\mu$ and with spread $\sigma$ in one dimension:
\begin{equation}
    R(x) = \frac{2}{\sqrt{3\sigma}\pi^{\frac{1}{4}}}\left(1-\left(\frac{x-\mu}{\sigma}\right)^2\right)\text{exp}\left(-\frac{(x-\mu)^2}{2\sigma^2}\right).
\end{equation}
If we write the argument of the exponent in terms of the index operator, we have
\begin{equation}
\begin{split}
    -\frac{(\hat{x}-\mu \hat{I}^{\otimes n_x})^2}{2\sigma^2}&= -\frac{1}{2\sigma^2}(\Delta x \hat{\imath} - \mu \hat{I}^{\otimes n_x})^2\\
    &= \sum_{m=1}^M c_m\hat{P}_m.
\end{split}
\end{equation}
In the second line we have noted that the first expression can be trivially evaluated in a symbolic manner, giving a collection of $M$ Pauli strings $P_m$ consisting of `$I$' and `$Z$' characters which are associated with Pauli operators $\hat{P}_m: \mathcal{H}_x\to \mathcal{H}_x$, summed over some coefficients $c_m\in\mathbb{C}$.

The operator whose diagonal in the computational basis encodes the gaussian part of this function is then
\begin{equation}\label{eq: Gaussian Op}
    \begin{split}
        \hat{g}(\hat{x}) &= \text{exp}\left(\sum_{m=1}^M c_m\hat{P}_m\right)\\
        &= \prod_m \left(\text{cosh}(c_m)\hat{I} + \text{sinh}(c_m)\hat{P}_m\right),
    \end{split}
\end{equation}
where in the second line we have used the fact that $[P_m,P_l]=0$ for all $m,\ l$. 

Similarly the polynomial prefactor can be rewritten as some other sum over $L$ Pauli operators, which we label with $l$, such that
\begin{equation*}
    \hat{R}(\hat{x}) = \frac{2}{\sqrt{3\sigma}\pi^{\frac{1}{4}}}\left(\sum_{l=1}^L c_l\hat{P}_l\right)\prod_m\left( \text{cosh}(c_m)\hat{I} + \text{sinh}(c_m)\hat{P}_m\right).
\end{equation*}

Clearly for such arbitrary functions the operator will not necessarily be unitary, and thus cannot be encoded in this manner onto a quantum computer. Nonetheless this provides a direct method for classically encoding arbitrary functions over an an exponentially large grid without ever having to store the function itself. Directly from the expression for Eq. \eqref{eq: Gaussian Op} for example we can construct each term in the product as an MPO, further contracting the MPOs to encode the gaussian function. Since the encoded function lives on the diagonal of the operator, to encode products of functions one just multiplies their MPO representations. Thus we can separately construct the MPO associated with the polynomial term, and multiply it with the Gaussian MPO. 

\begin{figure}
    \centering
    \includegraphics[width=\linewidth]{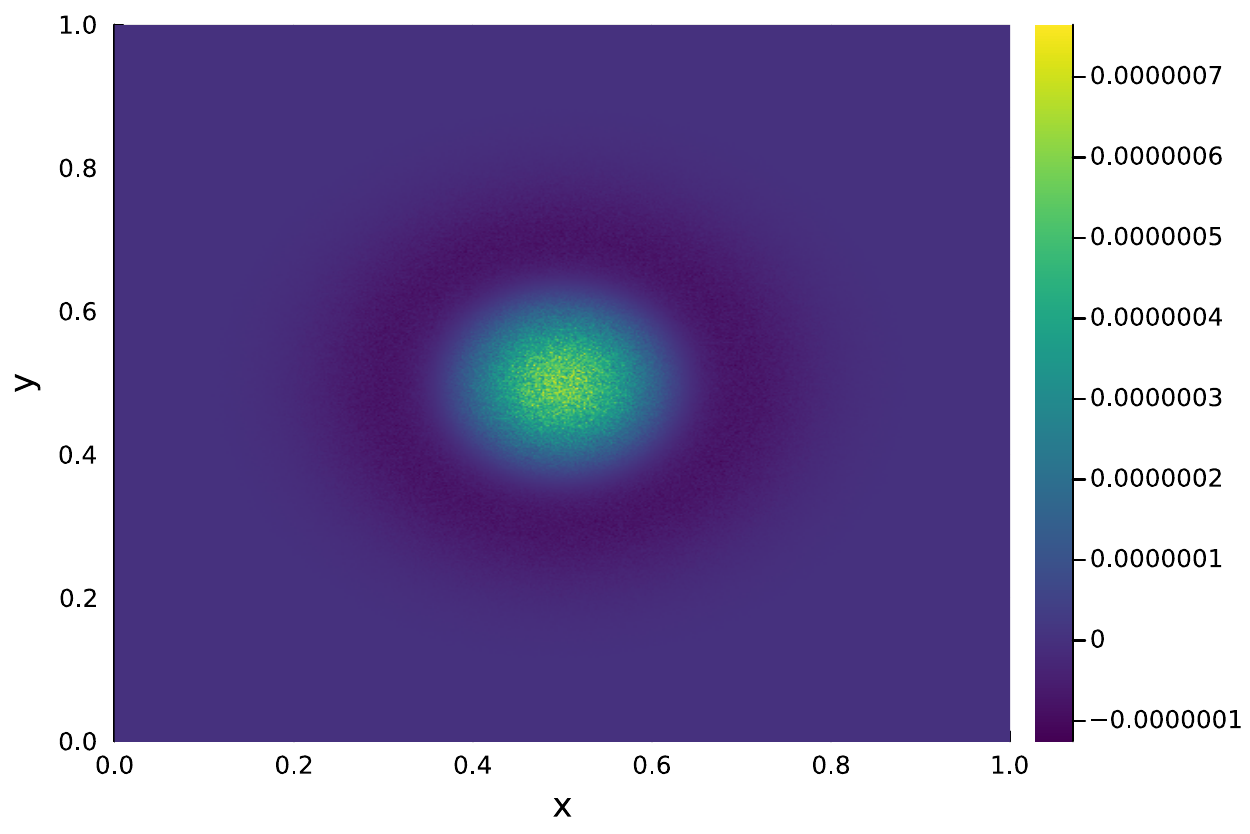}
    \caption{The two-dimensional Ricker Wavelett centered at $(0.5,0.5)$ with $\sigma=0.1$ represented across ${50}\times {50}$ qubit sites, or equivalently about $10^{30}$ grid points. This histogram is constructed from $10^7$ samples of the MPS state, grouped into $512\times 512$ bins. Preparing this state using ITensorMPS.jl ``out of the box'' with no attempt at optimization takes about 43 seconds on a laptop.}
    \label{fig:2d Rick}
\end{figure}

This process is straightforward to generalize to multidimensional functions, for example the two-dimensional Ricker Wavelett centered at $(\mu_x,\mu_y)$:
\begin{equation}\label{eq: 2d rick}
\begin{split}
    R(x,y) &= \frac{1}{\pi\sigma^4}\left(1-\frac{1}{2\sigma^2}\left((x-\mu_x)^2 + (y-\mu_y)^2\right)\right)\times\\
    &\text{exp}\left(-\frac{(x-\mu_x)^2}{2\sigma^2}\right)\text{exp}\left(-\frac{(y-\mu_y)^2}{2\sigma^2}\right).
\end{split}
\end{equation}

Here we simply recall that $\hat{x} \coloneqq\hat{x}\otimes\hat{I}_y = \Delta x\hat{\imath}\otimes\hat{I}_y$. In words: in any place where an isolated $\hat{x}$ appears, the corresponding Pauli on the $y$ register will be all identities and vice-versa. Obviously for directly correlated functions such as polynomials with cross terms like $\hat{x}^a\hat{y}^b$, there will be Pauli weight across both the $x$ and $y$ registers. In principle, expanding the products of the two Gaussians above will also generate crossed terms, but instead of evaluating them explicitly, we allow the MPO multiplication to build in the correlation. 

One immediate corollary of this method is that it becomes trivial to classically prepare MPS encodings of simple functions over exponentially large grids, so long as the bond dimension of the MPO representation is tractably small. First we prepare an even superposition across the computational basis through a simple application of Hadamard gates everywhere:
\begin{equation}\label{eq: all plus state}
    \ket{+}_x^{\otimes n_x}\ket{+}_y^{\otimes n_y}\ket{+}_z^{\otimes n_z} = \sum_{i,j,k}\frac{1}{\sqrt{2^{n_x+n_y+n_z}}}\ket{i}_x\ket{j}_y\ket{k}_z,
\end{equation}
where the sums over $x$, $y$, and $z$ grid points are indexed by $i$, $j$, and $k$ respectively, and run from $0$ to $2^{n_l}-1$ for each direction $l$. Then we construct the MPO which encodes the desired initial state $f(x,y,z)$ on its diagonal with an appropriate prefactor:
\begin{equation}\label{eq: f op}
    \hat{f}(\hat{x},\hat{y},\hat{z}) = \sqrt{2^{n_x+n_y+n_z}}\sum_{i,j,k}f(x_i,y_j,z_k)\ket{ijk}\bra{ijk}.
\end{equation}
One immediately sees that multiplying Eq. \eqref{eq: all plus state} into Eq. \eqref{eq: f op} will return the desired state:
\begin{equation}
\begin{split}
        \ket{f(x,y,z)} &= \hat{f}(\hat{x},\hat{y},\hat{z})\ket{+++}\\
        &=\sum_{i,j,k}f(x_i,y_j,z_k)\ket{ijk},
\end{split}
\end{equation}
which will of course \textit{not} be normalized. Normalization factors can be introduced explicitly into the multiplication, or the resultant MPS can be manually normalized if desired. However, care must be taken in the construction of these operators, as for large grid sizes the prefactors can easily lead to numerical instability. 

In Fig. \ref{fig:2d Rick} we demonstrate our approach by encoding the two-dimensional Ricker wavelett, Eq. \eqref{eq: 2d rick}, onto a $2^{50}\times 2^{50}$ grid, i.e. a grid with approximately $10^{30}$ grid points. Likely because the Ricker Wavelett is such a smooth, well-behaved function, the MPO encoding turns out to have a maximum bond dimension of 13 even with a Singular Value Decomposition (SVD) cutoff of $10^{-14}$. The resultant MPS of course has this same bond dimension. Building the MPOs and multiplying the MPS via straightforward and unoptimized usage of the Julia package ``ITensorMPS.jl'' \cite{itensor,itensor-r0.3} takes about 43 seconds on a laptop. Naturally the trade-off to having such a huge function in an MPS representation is that reconstruction of the wavefunction needs to be done via sampling the MPS. To do so, a random integer is drawn in the range of $\{0,\ldots,2^{n_x}-1\}$ for each dimension $l$, and the corresponding basis state is multiplied into the state MPS to get the value at that basis state, analogously to quantum state measurement. 

Often for applications the actual form of the wavefunction itself over the full space is less interesting than its shape over a small region, as in remote sensing \cite{Costa2019,Virieux2009}. In some cases what may be more relevant would be certain observables dependent on integrals over the wavefunction, which can be straightforwardly calculated using MPOs, assuming that the kernel of the integral in question is easy enough to encode in the MPO. Alternatively observables can be Monte Carlo integrated with samples from the MPS. The precise convergence properties of this sampling will depend on the observable in question.

Implementation of an MPO which consists of a large number of high-weight Paulis would be numerically challenging, particularly for MPOs whose Schmidt decomposition has a uniform distribution. More fundamentally, even constructing the Pauli strings necessary could be difficult in certain circumstances. There are $2^n$ possible combinations of `I' and `Z' characters across $n$ qubits, thus if there are very large sums which cannot be mapped to simpler products as in the case of the exponential function, this can quickly become unfeasible. Furthermore, as in the case of Eq. \eqref{eq: exact evolution}, operators like $|\hat{\mathbf{p}}| = \sqrt{\hat{D}_x^2 +\hat{D}_y^2+\hat{D}_z^2}$ would require setting up a system of equations allowing potentially all $2^n$ possible IZ Pauli strings to have non-zero weight in the square root, even before inserting this into further analytical functions like $\text{cos}(|\hat{\mathbf{p}}|t)$. 

\subsection{Conjugation by the Bit Reversal Operator}
Recall that the exponential operators in our evolution equation are conjugated by bit reversal operators $\hat{B}$ coming from the QFT. This turns out to be trivial to address in the Pauli string formulation, as the conjugation simply swaps the order of the characters in the Pauli string. That is for, say $\hat{P}_m = \hat{Z}\hat{Z}\hat{I}\hat{I}$,
\begin{equation}
    \hat{B}^{\dagger}(\hat{Z}\hat{Z}\hat{I}\hat{I})\hat{B} = \hat{I}\hat{I}\hat{Z}\hat{Z}.
\end{equation}
Thus before implementing any $\text{exp}(-i\Delta t\hat{\sigma}^l\hat{D}_l)$, we simply reverse each of the Paui strings in $\hat{D}_l$, and then apply Eq. \eqref{eq: Gaussian Op} with the Pauli strings $\hat{P}_m$ decorated by $\hat{\sigma}^l$ on the time qubit. 

\subsection{Small Angle Approximation: Quantum Algorithm}
For sufficiently well behaved functions it is straightforward to take Taylor expansions and insert the Pauli mapping of the index function into the truncated expansion. Although exploration of the convergence of such approaches is beyond the scope of this article, we note here that one can obtain a unitary approximation to Eq. \eqref{eq: evolution eq.} by performing such an expansion. In the small wave-vector limit we can take the first order expansion of $p_l\approx 2\pi k_l$. For a given direction this will look like:
\begin{equation}\label{DirPropagationCircuit}
     e^{-i\Delta t \hat{\sigma}^l \hat{D}_l} \approx  \sum_{k_l=-\frac{N_l}{2}}^{N_l/2-1}e^{-i2\pi \Delta t k_l \hat{\sigma}^l}|k_l\rangle\langle k_l|.
\end{equation}
This is the approach taken in the work done by Wright et. al in one dimension \cite{Wright2024}, although they arrive at this equation in a different manner. Due to having no non-commuting terms in one dimension, they were able to construct the exact evolution unitary circuit to arbitrary times. Inserting our index operator expansion into this equation produces a circuit very similar to theirs, although we take a different convention, namely that $\hat{k}_x = \hat{\imath} - \frac{N_x}{2}\hat{I}_x$. In this convention, the first $n_x$ qubit being in the $0$ state corresponds to the negative branch of the wavevectors. If desired one can simply take the unitary circuit described in detail in \cite{Wright2024} or our own Pauli string expansion, fix $t\to\Delta t$ and perform three-dimensional Trotterization with these unitaries acting on each spatial register. In this manner, one has a method to simulate the wave equation in the small wavenumber limit in three dimensions on a Quantum Computer. We will discuss the implications of this result in the conclusions, but for now we focus on the classical MPS algorithm.

\section{Results}
All calculations reported here were performed on a single workstation with 112 Intel Xeon Platinum 8280 CPUs set up across two NUMA nodes with 755 GB of RAM. 
\subsection{Two Dimensional Simulation}
\begin{figure}[h]
    \centering
    \includegraphics[width=\linewidth]{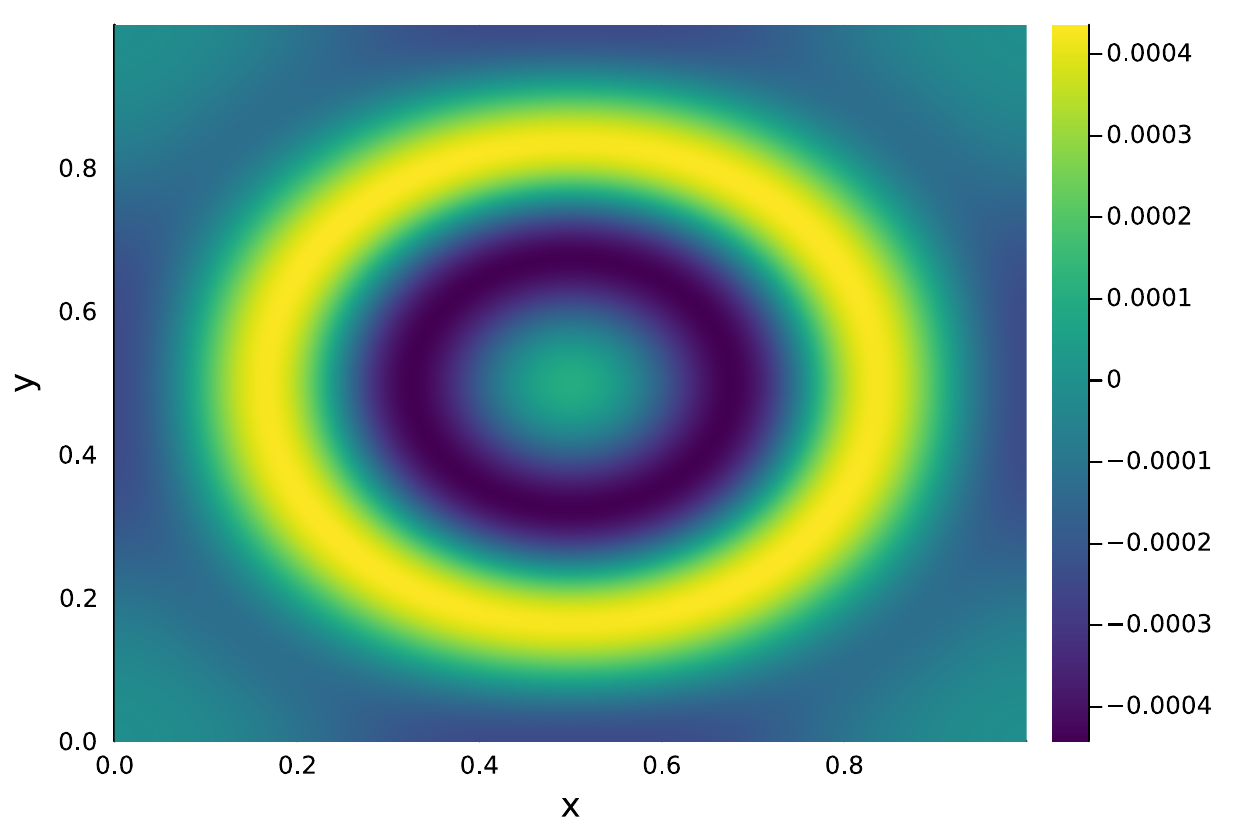}
    \caption{The two dimensional Ricker wavelett for $\mu=0.5$ and $\sigma=0.1$, evolved exactly to $t_f=0.3$ using Eq. \eqref{eq: exact evolution} on a $2^{12}\times 2^{12}$ grid. These results agree precisely with the direct RK4 integration of the real space equations of motion in Eq. \eqref{eq: direct EOM}. }
    \label{fig:exact R(t)}
    \centering
    \includegraphics[width=\linewidth]{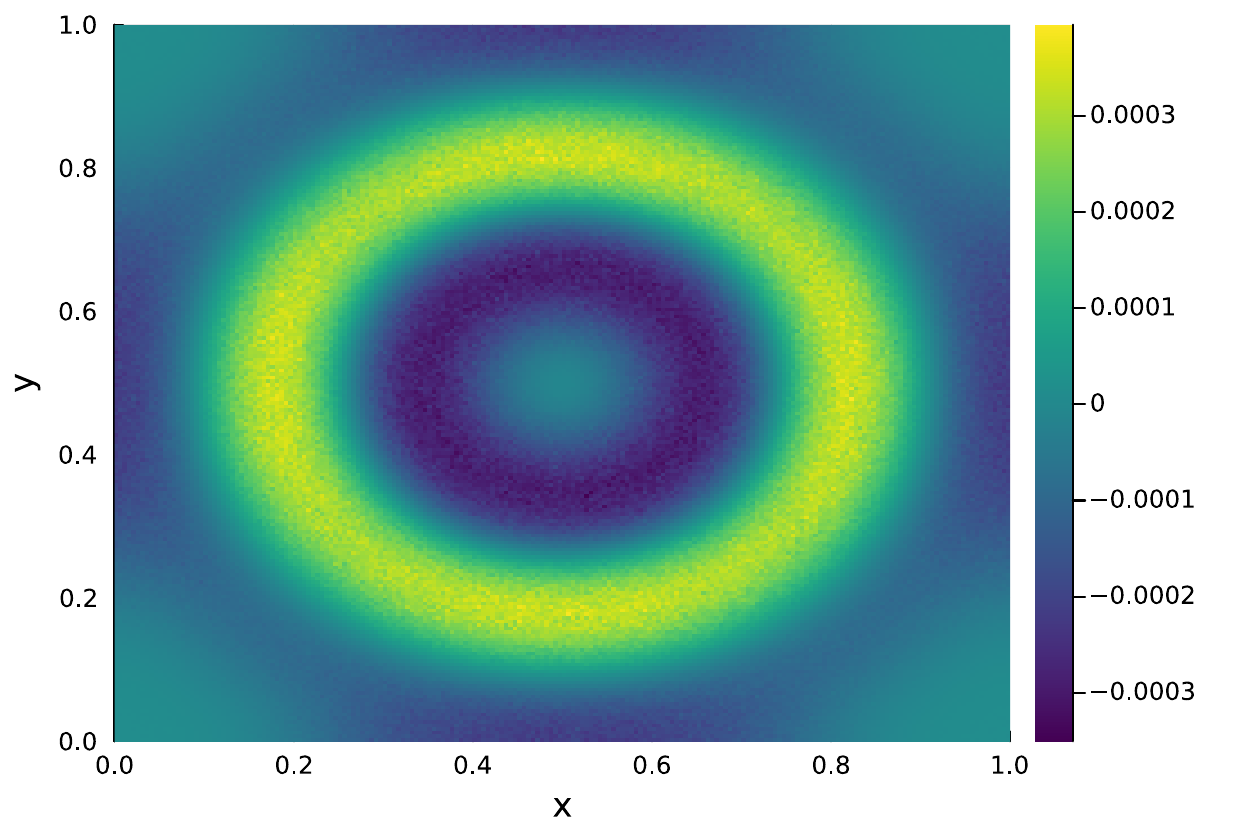}
    \caption{Two Dimensional Ricker wavelett for $\mu=0.5$ and $\sigma=0.1$, evolved using an MPS representation to $t_f=0.3$ using a Trotterization of Eq. \eqref{eq: evolution eq.} on a $2^{20}\times 2^{20}$ grid. This MPS is sampled across $10^7$ points which are sorted into $256\times 256$ bins. }
    \label{fig:MPS R(t)}
\end{figure}
The first step in our approach is to find a self-consistent set of initial conditions for our spinor wavefunction. From Eq. \eqref{eq: direct EOM}, we see that $(\partial_x-i\partial_y)\psi_1 = \partial_t\psi_0$. For validation of our algorithm, we simply start by taking $\partial_t\psi_0 = 0$ at the initial time, and thus for two dimensions, $\psi_1=0$ at time zero as well. Our initial condition is the Ricker wavelett with $\mu = 0.5$ and $\sigma=0.1$. We evolve to $t_f=0.3$, where the units of time are relative to the acoustic velocity $c=1$, and the box length, which we also set to 1. In Fig. \ref{fig:exact R(t)} we show the exact evolution of these initial conditions using Eq. \eqref{eq: exact evolution} on a $2^{12}\times 2^{12}$ grid. We find that these agree precisely with direct evolution of the real-space equations of motion, Eq. \eqref{eq: direct EOM} using Runge-Kutta 4 (RK4) propagation with $\Delta t = 0.0005$. The Fourier-space representation of the wavefunction and its back transformation to real-space are calculated via the FFT.

In Fig. \ref{fig:MPS R(t)} we show the Trotterized MPS solution for the same initial conditions with $20$ qubit sites in both the $x$ and $y$ registers corresponding to about $10^{12}$ total grid points. The MPO which encodes the initial spinor state is simply $\hat{U}_{\text{prep}} = \ket{0}_t\bra{0}\hat{R}(\hat{x},\hat{y}) + 0\times\ket{1}_t\bra{1}\hat{I}_{x,y}$, where $\ket{0}_t\bra{0}$ is the $0$ state projector on the time qubit, and we explicitly force the $1$ state projector term to be zero such that $\ket{\Psi}\to(\psi_0, 0)$. Here we use $\Delta t = 0.0005$, and set the SVD cutoff used in the MPO/MPS zip-up multiplication algorithm to $10^{-14}$. In constructing the Trotterization operators in each direction, we also enforce an SVD cutoff of $10^{-14}$, and we only consider coefficients in the Pauli strings whose absolute value is greater than $10^{-8}$. We utilize Eq. \eqref{eq: Gaussian Op} to construct the $\hat{D}_l$ operators, since their entries are $p_l = N_l\text{sin}(\frac{2\pi}{N_l}k_l)$. We then use Eq. \eqref{eq: Gaussian Op} again to encode the time evolution operators $\text{exp}(-i\Delta_t\hat{\sigma}^l\hat{D}_l)$ after symbolic conjugation by the bit reversal operator across the spatial registers. In practice we find that the maximum bond dimension attained in this propagation is about $25-30$ depending on system size. The coarseness in the plot is a function of the density of sampling points and the binning of the histogram.
\begin{figure}
    \centering
    \includegraphics[width=\linewidth]{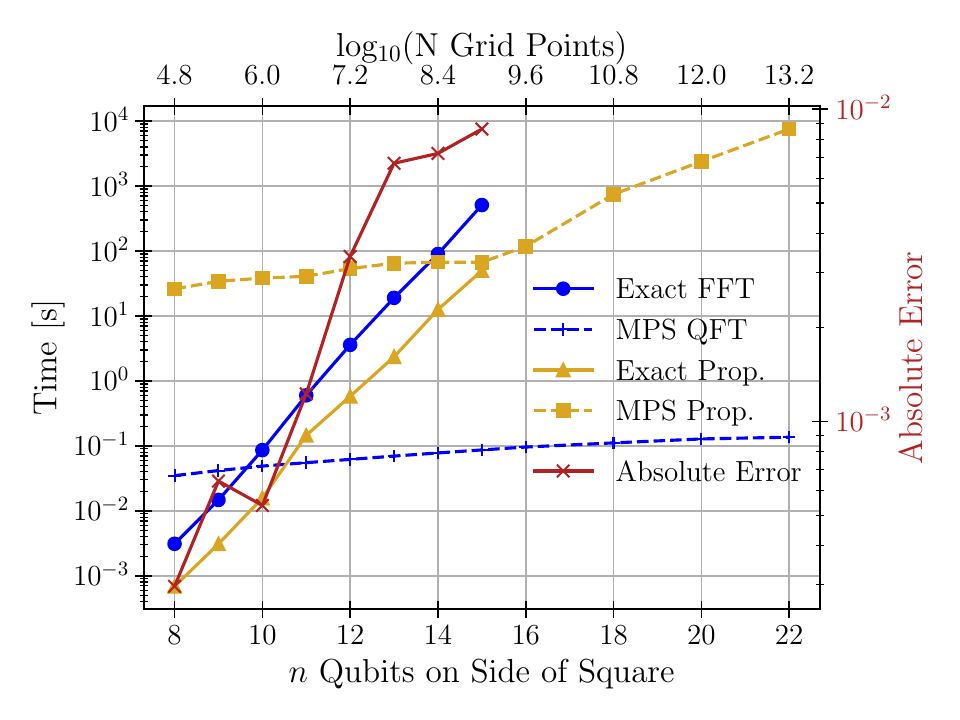}
    \caption{Scaling behavior of the algorithm in two dimensions for a Ricker wavelett initial condition. The left y-axis shows the time associated with each operation named in the first four entries of the legend, where `Prop.' refers to the time for propagating the Exact or MPS solution, equations \ref{eq: exact evolution} and \ref{eq: Trotter MPS Evol} respectively, and FFT/QFT refer to the Fast Fourier Transform and Quantum Fourier Transform algorithms, respectively. The right y-axis shows the absolute error between the MPS and exact solution: $|\psi_0^{\text{MPS}}-\psi_0^{\text{exact}}|$ with a maximum value of $0.86\%$. The bottom x-axis shows the number of qubits for a given spatial direction. The top x-axis shows the log base 10 of corresponding total number of grid points, i.e. $\text{log}_{10}(2^{2n})$. Past $n=15$ on a side we no longer include the exact calculation. The MPS algorithm begins to be faster than direct representation on the grid for grids of size $2^{14}$ on a side due to the FFT versus QFT scaling, although alternative MPS propagation strategies could potentially improve this. See the text for details.}
    \label{fig:rick scaling}
\end{figure}

In Fig. \ref{fig:rick scaling} we analyze the scaling behavior of the algorithm in two dimensions. On the left y-axis we show the time required for the FFT, the QFT, the exact propagation, and the MPS propagation. The exact evolution of $\psi_0$ is calculated using Eq. \eqref{eq: exact evolution} in a single element-wise vector multiplication. Since it is not used at the final time, $\psi_1$ is not evolved. The time to construct the functions in Eq. \eqref{eq: exact evolution} is not included. As before the MPS propagation is done to $t_f=0.3$ for $\Delta t=0.0005$ meaning $N_t=600$ time steps, with the same SVD cutoff and Pauli string coefficient cutoff. The time to construct the MPO responsible for MPS state-preparation as well as the Trotterized evolution MPOs are not included, as they can be pre-calculated in practice. On the right y-axis we show the absolute error, $|\psi_0^{\text{MPS}}-\psi_0^{\text{exact}}|$, for the grids where we were able to represent the exact solution. 

The time for the FFT is calculated as twice the time required to transform the initial state $\psi_0$, since one would generally also need to transform $\psi_1$, plus the time to back-transform only $\psi_0$, as $\psi_1$ is not of interest. The FFT is performed on the two axes of a matrix representation of $\psi_{0}(x,y)$. The scaling of the FFT roughly conforms to the theoretical $\mathcal{O}(n^2\text{log}(n^2))$ expression with a prefactor in this case of about $10^{-6.5}$. Extrapolating this scaling well beyond the physical memory limits of the device to the $n_l=22$ on a side case, the FFT alone would take about 2.3 years. The scaling of the QFT in qubit size, $n$, is consistent with the results from Chen, Stoudenmire and White \cite{Chen2023}. Due to superposition, the QFT is transforming both $\psi_0$ and $\psi_1$ simultaneously. Purely because of the performance of QFT over FFT, a time advantage is gained in the MPS calculation at grids with $2^{14}$ grid points on a side. The absolute error, while less than $1\%$ for the exact solutions considered, can be reduced by decreasing the time-step, or in principle through more advanced time evolution algorithms~\cite{Haegeman2016prb,Orus2019,Kohler2019}. Beyond the scale where exact evolution is possible, self-consistent convergence must be considered with respect to SVD cutoff, time-step and Pauli string coefficients. 

\subsection{Three Dimensional Scaling}
\begin{figure}
    \centering
    \includegraphics[width=\linewidth]{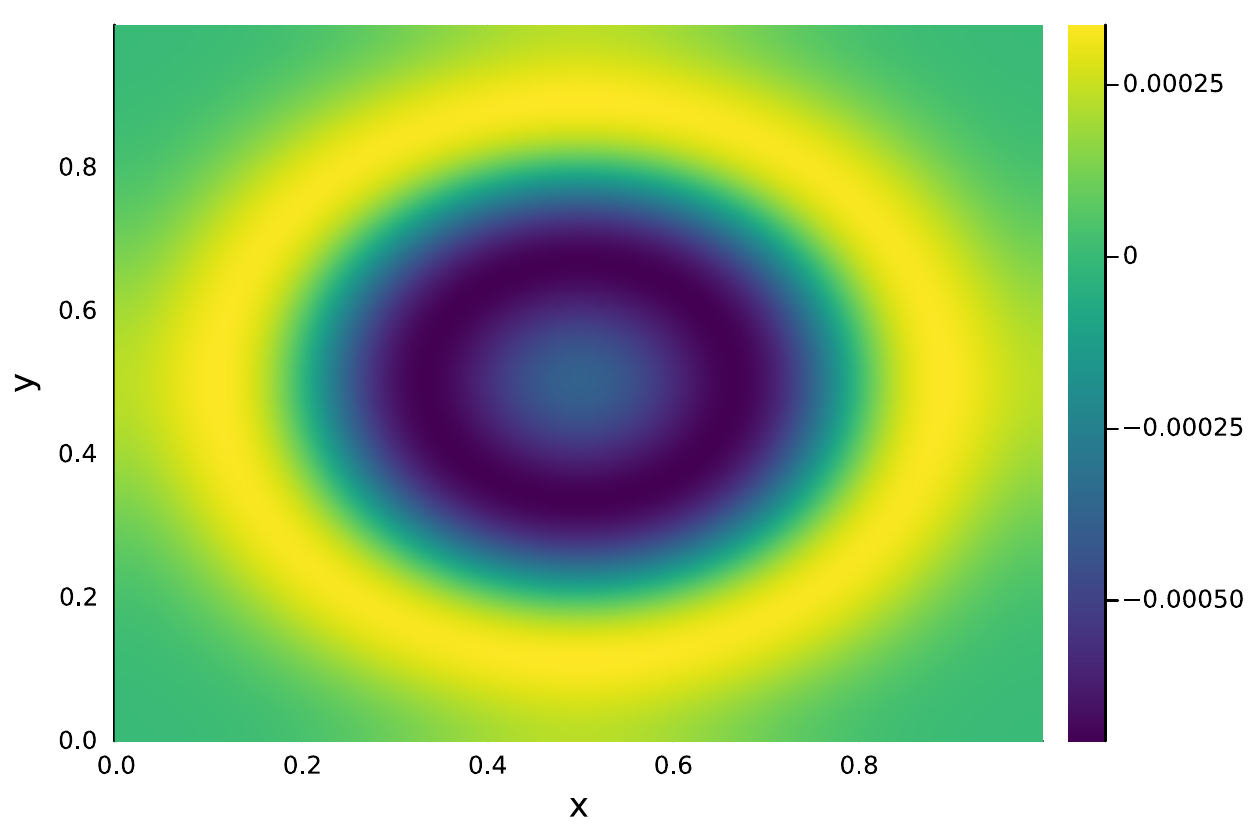}
    \caption{The tapered three dimensional Gaussian function for $\mu=0.5$ and $\sigma=0.1$, evolved exactly to $t_f=0.3$ using Eq. \eqref{eq: exact evolution} on a $2^{8}\times 2^{8}\times2^{8}$ grid, sliced through the $z=0$, $xy$ plane. At the $z=0$ plane we expect the errors due to the approximate initialization of $\psi_1$ to be minimal as $\partial_z\psi_0=0$ at $t=0$ on this plane. These results agree precisely with the direct RK4 integration of the real space equations of motion in Eq. \eqref{eq: direct EOM}. }
    \label{fig:exact G(t)}
\end{figure}

In three dimensions the first task for the algorithm is to determine the non-trivial initial conditions. We analyze the case of $\psi_0=g(x,y,z)$, where $g$ is the product of standard Gaussian functions in each direction with $\sigma=0.1$ and $\mu = 0.5$, and the initial velocity $\partial_t\psi_0 = 0$. We can assume a separable form for $\psi_1 = f(x,y)\partial_zg(z)$, where $f(x,y) = a(x,y) + i b(x,y)$. Since $\text{Im}(\psi_0) = 0$, the initial conditions on $\psi_1$ must satisfy the following system of differential equations:
\begin{equation}
\begin{split}
        \partial_ya(x,y)- \partial_xb(x,y)  &=0 \\
        \partial_xa(x,y) + \partial_y b(x,y) &= g(x,y).
\end{split}
\end{equation}
By assuming that $a = \partial_x \Lambda(x,y)$ and $b = \partial_y \Lambda(x,y)$ for some unknown function $\Lambda(x,y)$, we trivially satisfy the first equation and the second equation becomes
\begin{equation}\label{eq:Poisson}
    \begin{split}
        (\partial_x^2+\partial_y^2)\Lambda(x,y) = g(x,y).
    \end{split}
\end{equation}
This Poisson equation can be solved under periodic boundary conditions using standard numerical techniques. However, for the sake of analytical insight—and with the hope of later devising a way to encode this as an MPO—we consider an approximate solution.
It turns out that we can approximately satisfy the Poisson equation under \textit{open boundary conditions} with the following functional form:
\begin{equation}
    \begin{split}
        \psi_1 &\approx \sigma^2\frac{w}{|w|^2}\left(e^{-\frac{|w|^2}{2\sigma^2}} -1\right)\\
        w &= (x-\mu) + i(y-\mu)
    \end{split}
\end{equation}
Note this is only an approximate solution because the first term in the first equation is non-analytic at $x=y=\mu$. The second term regularizes the derivative in the region of $(\mu, \mu)$, and the initial conditions are satisfied within a small error throughout the plane. However this `solution' is only valid away from the boundaries, since for our periodic boundary conditions, this breaks down at the edge. For the sake of considering how the MPS algorithm scales in three dimensions, we arbitrarily impose a restriction on the initial waveform by multiplying it with a modified `power-of-sine' window function $m(x)$, which has a wide plateau approximately equal to 1 around the center of the Gaussians and smoothly tapers to zero at the boundaries:
\begin{equation}\label{eq: tapered Gaussian}
\begin{split}
    \psi_0 &= \frac{m(x)m(y)}{N}e^{-\frac{1}{2\sigma^2}(x-\mu)^2}e^{-\frac{1}{2\sigma^2}(y-\mu)^2}e^{-\frac{1}{2\sigma^2}(z-\mu)^2}\\
    m(t) &= \text{cos}(\pi(t-\mu))^{|\pi (t-\mu)|/\alpha}\\
    \psi_1 &\to \frac{m(x)m(y)}{N}\psi_1
\end{split}
\end{equation}
$N$ is a normalization condition such that $|\psi_0|^2+|\psi_1|^2 = 1$, and we set our tapering factor $\alpha=1.2$. The difference between this function and a normal product gaussian is very small for sufficiently small $\sigma$ and large $\alpha$. We force $|\psi_0| = |\psi_1| = \frac{1}{\sqrt{2}}$. Encoding both the window function and the $1/|w|^2$ factor in $\psi_1$ directly as MPOs is non-trivial with our approach, so for the MPS simulation we simply construct these two functions and then convert them to MPS forms. Furthermore it means that our simulation is only valid for short times, before the error from approximating $\psi_1$, and the inconsistency at the boundaries accumulates. However, we include these results with this caveat firmly in mind, not so much for their real-world accuracy which could be improved in many ways, but for the sake of understanding the scaling behavior of the MPS algorithm in three dimensions. In Fig. \ref{fig:exact G(t)} we show what the tapered three dimensional Gaussian wavefunction looks like on the $xy$ plane at $z=0$ after being evolved to $t_f=0.3$. This agrees very closely with what Gaussian evolution in two dimensions looks like, because in this plane $\psi_1 = 0$ at time zero due to being the node in $\partial_zg(z)$. 

\begin{figure}
    \centering
    \includegraphics[width=\linewidth]{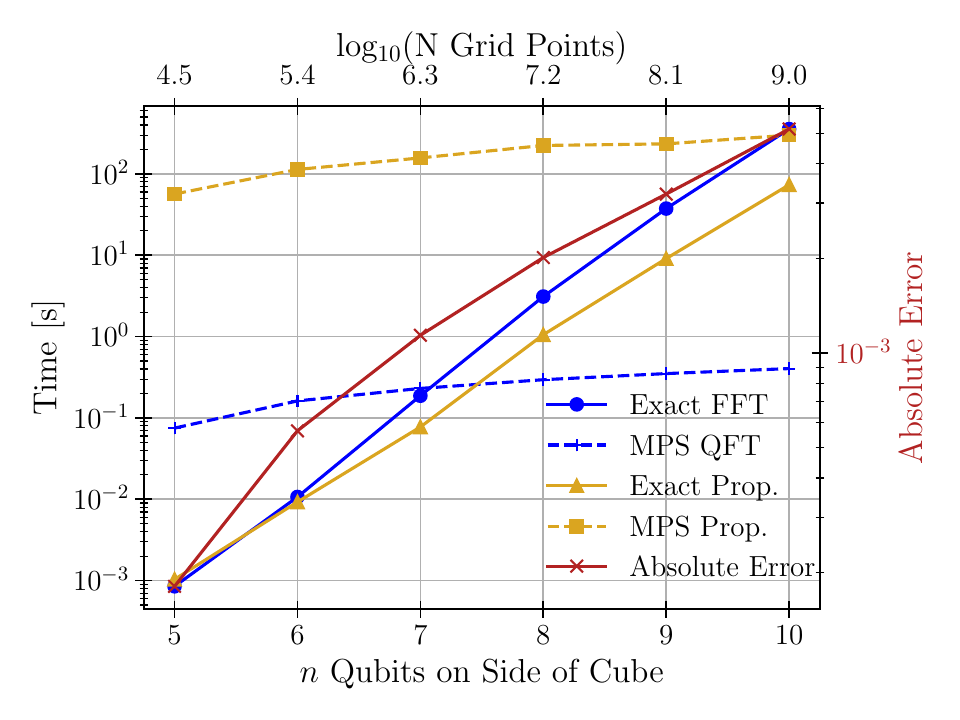}
    \caption{Scaling behavior of the algorithm in three dimensions for a tapered Gaussian initial condition, Eq. \eqref{eq: tapered Gaussian}, evolved to $t_f=0.3$ for $\Delta t=0.0005$. The left y-axis shows the time associated with each operation named in the first four entries of the legend, where `Prop.' refers to the time for propagating the Exact or MPS solution, and FFT/QFT refer to the Fast Fourier Transform and Quantum Fourier Transform respectively. The right y-axis shows the absolute error between the MPS and exact solution: $|\psi_0^{\text{MPS}}-\psi_0^{\text{exact}}|$ with a maximum value of $0.5\%$. The bottom x-axis shows the number of qubits responsible for a given spatial direction. The top x-axis shows the log base 10 of corresponding total number of grid points, i.e. $\text{log}_{10}(2^{3n})$. Time advantage of the MPS is gained at grids of $2^{10}$ grid points on a side}
    \label{fig:Gaussian scaling}
\end{figure}

In Fig. \ref{fig:Gaussian scaling} we show the scaling behavior of the simulation of our tapered Gaussian in three dimensions. As before, we evolve to $t_f=0.3$ with $\Delta t = 0.0005$, MPS/MPO SVD cutoff set to $10^{-14}$ and Pauli string coefficient cutoff at $10^{-8}$. At $n=10$ on a side the initial maximum bond dimension of the MPS state is 30, and the maximum bond dimension at the end of that simulation is 51. Again, due to the performance of the QFT over the FFT, the MPS algorithm ourperforms the direct evolution - in this case at $2^{30}$ grid points - with an absolute error compared to the exact evolution of $0.5\%$. Extrapolation of the FFT versus MPS propagation time again shows that, if appropriate initial conditions can be encoded as MPSs for particular boundary conditions, that the potential gains in computation time would be substantial. 

\section{Discussion and Conclusion}
We have derived a first-order in time propagation method for the isotropic acoustic wave equation by analogy with the massless Dirac equation. We then applied a uniform real-space discretization under periodic boundary conditions to derive the associated unitary quantum circuit. Using this formulation, we obtained an exact solution for arbitrary time evolution in Fourier space, which was benchmarked against evolution in real space using RK4. We then presented our approach for encoding simple analytic functions into MPSs and constructing diagonal MPOs~\cite{Haegeman2017} with simple functions on their diagonals. With this method we were able to encode the two-dimensional Ricker wavelett onto a 100 site MPS, equivalent to about $10^{30}$ grid points. By sampling this MPS we are able to reconstruct the initial state in a controlled manner and calculate properties of this wavefunction across particular sections of the grid. This method can be used to simulate the IAW across two and three dimensions in the small wave-vector limit on a quantum computer; in the one dimensional limit, our approach reduces to the method of Wright et. al.~\cite{Wright2024}.

Using these results we assessed the accuracy and scaling of the MPS encoding of the IAW for two and three dimensions using the ITensorMPS.jl Julia package \cite{itensor,itensor-r0.3}. The initial conditions for our algorithm are trivial when the initial velocity is zero, allowing us to evolve the two-dimensional Ricker wavelett on a grid of about $10^{13}$ points. Due to the small entanglement properties of the core of the MPO representation of the QFT \cite{Chen2023}, and the efficient symbolic assessment of the otherwise high entanglement bit reversal operation, our calculation gained significant speedups over direct evolution on the grid. Extrapolating our results on a modestly sized workstation to the $10^{13}$ grid point case, even if the hardware had the required 160 TB required to hold the initial conditions in memory, the FFT alone would require 2.3 years of calculation time, while the MPS algorithm takes about two hours and five minutes. We found that with an MPS/MPO cutoff of $10^{-14}$ and a timestep of $\Delta t= 0.0005$ we could keep the absolute error of the MPS calculation to $0.86\%$ on grid sizes of around $10^{8.4}$. 

In the three dimensional case, we found that writing down an analytical expression for the MPS encoding of the initial conditions is not straightforward. Nonetheless we made an approximation, whose long-time accuracy and accuracy for our boundary conditions breaks down. We used this approximation for the sake of assessing the scaling behavior of our algorithm. This required directly representing the initial conditions on the grid and then encoding them to MPSs, limiting the size we could simulate. Given this caveat, we still found that as expected the MPS algorithm will outperform the direct simulation in calculation time starting in this case at about $10^9$ grid points, again mostly due to the scaling of the QFT. 

Moving forward we see many possible implications and follow up studies to build on this work. Most obviously, finding proper initialization methods for the three dimensional case, for example by solving Eq. \eqref{eq:Poisson} using standard numerical techniques, or by finding it directly in an MPS form. Generalizing this initialization routine is crucial for scaling to arbitrary applications. Another area with room for improvement is the contraction order of the MPO propagators, where applying more advanced contraction techniques, say in a term-by-term assessment of the optimal application of Eq. \eqref{eq: Gaussian Op} to the MPS, could yield significant performance gain in the time required to do MPS propagation. Furthermore, looking at the structure of our Trotterized evolution in Fig. \ref{fig:Prop full}, we see that a one-dimensional Tensor Train geometry might not be the most efficient way to evaluate Eq. \eqref{eq: Trotter MPS Evol}. Exploring alternative tensor geometries could also result in considerable performance gain. We note that since the MPS equation of motion for the small angle limit can be evolved in a controlled manner to large numbers of qubits, this provides a robust benchmark against which the analogous quantum circuit can be tested for either error mitigation method testing \cite{Cai2023,Lively2024,Mangini2024} or tests for quantum utility or even advantage.

Finally we conclude that the approach presented in this work could be extended to other partial differential equations or general boundary conditions \cite{Costa2019}. For example, potential applications beyond the acoustic wave equation could include coupled oscillator equations \cite{Babbush2023} or Maxwell's equations \cite{Vahala2020,Koukoutsis2023,Koukoutsis2023App,Koukoutsis2024}. In principle, the MPS algorithm can also be straightforwardly applied to other first-order in time PDEs such as the advection equation \cite{Brearley2024}. For other second order in time differential equations, it is interesting to consider whether generalizations of the Dirac operator, known as half-iterates, could be constructed in analogy to our method here. We expect that the continuing development of robust methods of Tensor Network encoding of PDEs and their initial conditions will provide substantial near term computational advantage, while simultaneously advancing Quantum Computational algorithms which may some day surpass them.  

\bibliographystyle{IEEEtran}
\bibliography{bibliography.bib}

\end{document}